# The relative impact of private research on scientific advancement

Giovanni Abramo*
*Laboratory for Studies in Research Evaluation*
*at the Institute for System Analysis and Computer Science (IASI-CNR)*
*National Research Council of Italy*
ADDRESS: Istituto di Analisi dei Sistemi e Informatica, Consiglio Nazionale delle Ricerche,
Via dei Taurini 19, 00185 Roma - ITALY
giovanni.abramo@uniroma2.it

Ciriaco Andrea D'Angelo
*University of Rome "Tor Vergata" - Italy and*
*Laboratory for Studies in Research Evaluation (IASI-CNR)*
ADDRESS: Dipartimento di Ingegneria dell'Impresa, Università degli Studi di Roma "Tor Vergata", Via del Politecnico 1, 00133 Roma - ITALY
dangelo@dii.uniroma2.it

Flavia Di Costa
*Research Value s.r.l.*
ADDRESS: Research Value s.r.l., Via Michelangelo Tilli 39, 00156 Roma- ITALY
flavia.dicosta@gmail.com

**Abstract**
Literature about the scholarly impact of scientific research offers very few contributions on private sector research, and the comparison with public sector. In this work, we try to fill this gap examining the citation-based impact of Italian 2010-2017 publications distinguishing authorship by the private sector from the public sector. In particular, we investigate the relation between different forms of collaboration and impact: how intra-sector private publications compare to public, and how private-public joint publications compare to intra-sector extramural collaborations. Finally, we assess the different effect of international collaboration on private and public research impact, and whether there occur differences across research fields.

**Keywords**
*Industry; Research collaboration; Scientific impact; Bibliometrics; Italy*

**Acknowledgment**
We wish to thank Francesca Apponi for her contribution to organizations' name disambiguation.

___________________
*\* corresponding author*

# 1. Introduction

This work tries to assess the impact of Italian private research on scientific advancement, with a particular focus on private-public research collaboration, whose results represent the largest share of private-sector publications.

Research-based innovation is a source of competitive advantage in an increasing number of economic sectors. In most OECD countries, private R&D expenditures are higher than public (OECD, 2020a). In order to appropriate the benefits of their R&D investments, private corporations remain very wary of making their R&D results public. They often recur to industrial property protection to prevent competitors from benefiting of knowledge spillovers. On the contrary, universities and public research organizations (U&PROs) tend to encode their research results in written form and favor their rapid and spread diffusion, responding to incentive systems based on the evaluation of the impact of their research production. As a result, in every country we witness the private sector holding a disproportionate share of the country's overall intellectual property. Conversely, we have U&PROs authoring the larger share of a country's publications.

While there is ample literature on the assessment of the scholarly impact of U&PROs' research, studies on private research tend to focus on research impact on technological and economic development (the former generally hold publications in scientific journals as the basic unit of analysis, while the latter patents). In theory, there could be two reasons for paucity of studies on the scholarly impact of private research. One is that such investigation is considered as not worth it, because the share of scientific research vis-à-vis applied research and technological development by the private sector is negligible. The other reason is that a critical step of the scholarly impact assessment process requires manual identification and name reconciliation of all organizations authoring publications, to distinguish private sector from public research. For large-scale studies, this reveals a formidable task, and requires a profound knowledge of the research systems and languages of countries under observation. We propend for the latter reason. This work intends to contribute to fill in the gap, trying to answer a number of questions on the scientific impact of private research.

The first question that comes to mind is: What is the average impact on scientific advancement of private-sector publications as compared to public-sector ones? And, strictly related to that: What is the share of highly cited publications by the private sector as compared to the public sector?

The increasing complexity and cost of research to solve global societal problems (the current COVID-19 pandemic emergency is a case in point), mostly interdisciplinary in nature, and ad hoc incentive systems and science policies and initiatives towards collaborative research, has led scientists to increasingly rely on research collaboration (Wuchty, Jones, & Uzzi, 2007; Larivière, Gingras, Sugimoto, & Tsou, 2015). Furthermore, since public and private organizations are stimulated to collaborate more and more to promote knowledge transfer not only ex post, but already in its production steps (Perkmann et al., 2013; Etzkowitz & Leydesdorff, 2000), what one finds is that most publications by private (public) organizations are actually fruit of collaboration, within the private (public) sector or with the public (private) sector.

Another question then is: How do different forms of collaboration relate to the impact of private-sector publications? Are there differences from those of their public counterparts? And how do private-public joint publications compare to other types of extramural collaborations? Looking at collaboration from the perspective of the public scientist, subject to assessment based on the impact of its scientific production, all other motivations being equal, is she worth cooperating with colleagues in the private sector?

Another phenomenon that has been taking place for decades, but had a burst with the advent of the internet, is that research activities are increasingly globalizing. The rate of international research collaboration, as shown in the authorships of papers, is increasing (National Science Board, 2020). Because it is known that international publications have higher scientific impact (Goldfinch, Dale & DeRouen Jr, 2003; Schmoch & Schubert, 2008; Abramo, D'Angelo, & Solazzi, 2011), another question follows: in presence of co-authors affiliated to foreign institutions, is there an impact increase for private domestic publications similar to that of their public counterparts? Finally, we investigate whether there occur differences in the above, across research fields.

To the best of our knowledge, two other studies only tried to assess the scientific impact of private research, in particular of private-public research collaborations. The first one by Lebeau, Laframboise, Larivière, and Gingras (2008) shows that Canadian publications fruit of private-public research collaborations are generally published in less prestigious journals (i.e. with a lower impact factor, IF), but are on average more cited than those by university only, in the same field and year of publication (Lebeau, Laframboise, Larivière, & Gingras, 2008). Unfortunately, the analysis is purely descriptive, as the authors do not control for such covariates as the number and nationality of co-authors and type of publication. It is known in fact that citation rates are positively correlated with the number of authors (Persson, Glänzel, & Danell, 2004; Wuchty, Jones, & Uzzi, 2007; Larivière, Gingras, Sugimoto, & Tsou, 2015), and it is so for Italian publications as well (Abramo & D'Angelo, 2015). Furthermore, internationally co-authored publications are more highly cited than domestic (van Raan, 1998; Adams, 2013; Kumar, Rohani, & Ratnavelu, 2014), and it is definitely so in Italy as well (Abramo, D'Angelo, & Murgia, 2017), even when including Italian self-citations (Abramo, D'Angelo, & Di Costa, 2020). Finally, citation rates vary across document types, i.e. articles, reviews, conference proceedings, etc. (Tahamtan, Safipour Afshar, & Ahamdzadeh, 2016).

Private research activities stress that the knowledge is kept in-house, and pursue its commercial exploitation. Therefore, it is expected that, as compared to U&PROs, private corporations tend to engage less in international and multi-institution collaborations, and the share of their papers published in conference proceedings is larger than that of U&PROs, while the share of reviews is lower. When comparing then the publication impact of private vs public research, one needs to control for such variables as the above ones.

The other study by Bloch, Ryan, and Andersen (2019), partly overcomes the above limits, conducting an inferential analysis with reference to the Danish case. The authors show that there are no significant differences in citation impact between public-private joint publications and publications authored by exclusively public authors. Rather, among publications resulting from international collaboration, there is a higher citation impact for public-private publications than for public-only publications.

In this work, we will evaluate the impact of the scientific production of the private sector, in its different types (single- and multi-authored, fruit of intra- or extramural collaboration, domestic or international, with other private and/or public organizations), comparing it with that of the corresponding publications by U&PROs.

We will adopt a bibliometric approach, based on the analysis of Italian scientific publications indexed in Web of Science (WoS) in the period 2010-2017. The authors' affiliations will be analyzed to distinguish private from publicly authored publications. The restriction of the analysis to our home country only is due to the formidable task of manually disambiguating and unifying U&PROs and private organizations names, which requires profound knowledge of the context and language. Each publication will be assigned to the WoS subject category (SC) of the hosting journal, to carry out the analysis also at field level.

The work is structured as follows. The next section presents a literature review on private-public research collaboration motivations and effects. The third section deals with the theoretical base for the assessment of the scholarly impact of research The fourth section presents the dataset used and describes the methodology adopted. In the fifth section, the results of the analysis are shown and commented. Finally, the sixth concludes the work discussing the relevant findings and their implications for research policies.

## 2. Private-public research collaboration: motivations and effects

Most of private R&D leading to results encoded in publications is conducted in collaboration with U&PROs. That partly explains why most of what we know on the scholarly impact of private research, actually concerns private-public research collaborations.

Similarly to other countries, 90 percent of overall 2010-2017 Italian publications are fruit of research collaboration: 96 percent in the case of private sector publications. Furthermore, 87 percent of private sector publications are fruit of extramural collaborations: 70 percent with domestic U&PROs.

Joint research between U&PROs and industry produces effects that positively influence economic growth, social value, and competitiveness. In recent years, policy makers have therefore increased their actions to promote public-private collaboration by acting on both partners through: i) research policies aiming at promoting the so-called third mission of universities (protection, licensing and commercialisation of research results and entrepreneurship in general); ii) innovation policies to encourage companies to interact with public research organisations (Perkmann et al., 2013; Guimón & Paunov, 2019).

Beyond the incentive schemes, the motivations underlying public-private cooperation are different for the two partners. The private sector has an interest in collaborating with the public to have access to skills with which to create new knowledge to exploit for commercial purposes (Bekkers & Freitas, 2008; Perkmann et al., 2013).

The public researcher benefits from accessing instrumental assets not in his or her possession, but accesses also economic and financial benefits of various kinds (Garcia, Araújo, Mascarini, Santos, & Costa, 2020). However, public-private collaboration remains an important means by which public and industrial researchers can share knowledge in synergy (Cohen, Nelson, & Walsh, 2002; Mowery & Sampat, 2004).

Public-private collaboration also involves costs, so the intensity of the collaboration depends on the balance between costs and benefits arising from these relationships. The pressure from national research evaluation exercises, undertaken in an increasing number of countries, certainly leads public researchers to pay particular attention when selecting research projects to be included in their agenda, due to the probability of reaching outcomes (and resulting publications) with a high impact on the international community.

Empirical evidence on the effects of public-private partnerships on scientific performance is not fully aligned.

Many scholars suggest that academics and PROs' scientists collaborating with private companies may experience lower publication rates because of the industry's restrictions due to intellectual property protection through confidentiality requirements (Louis, Jones, Anderson, Blumenthal, & Campbell, 2001; Thursby & Thursby, 2002; Evans, 2010; Shibayama, Walsh, & Baba, 2012). Even the mere financial support and sponsorship by industry increases the expected probability of publication delay, as compared to other forms of funding (Czarnitzki, Grimpe, & Toole, 2014). Patenting a scientific discovery resulting from a public-private joint project, may decrease the probability for a scholar to be involved in follow-on research based on that discovery (Murray & Stern 2007, Williams, 2013). Overall, therefore, it seems that academics and PROS's scientists collaborating with industry are likely to find it more difficult on average to publish the results of their research.

In contrast, Garcia, Araújo, Mascarini, Santos, and Costa (2020) in their recent study on university-industry scientific collaborations in Brazil in the period 2002-2008 provide empirical evidence that long-term collaborative research projects positively influence the scientific output of the involved groups as compared to that of non-cooperating control groups. The long-term commitment clearly strengthens trust and reduces the barriers related to conflicts over intellectual property and contract management. A similar conclusion is reached by Bikard, Vakili, & Teodoridis (2019), who claim that academics collaborating with the industry produce more publications and fewer patents than their colleagues who do not.

Zhang and Wang (2017) studied the influence of university-industry collaboration on the research performance of a set of 804 academic researchers in engineering using the social networking framework. The results show that the intensity of collaboration has a negative effect on the research performance of academics (h-index), while relational social capital, measured by the strength of the link, has a positive moderating effect.

Banal-Estañol, Jofre-Bonet, & Lawson (2015) have studied the effect of university-industry collaboration on the results of academic research in the UK. They used a longitudinal dataset covering the period 1986-2007 containing the individual features, publications, research funds, and patents of all researchers affiliated to engineering departments of 40 major UK universities. The results show that productivity (simply measured by the number of publications) increases with the intensity of collaboration, but only up to a certain threshold, beyond which the effect becomes negative. The average quality of publications, on the other hand, is not significantly affected by collaboration with industry.

Abramo, D'Angelo, Di Costa, and Solazzi (2009) have shown that Italian academics who collaborate with the private sector register research performance that is superior to that of colleagues who are not involved in such collaboration. A subsequent study has shown the importance of geographic proximity in companies' choices of university partner, and that in a large proportion of cases private corporations could have chosen more qualified research partners in universities located closer to the place of business (Abramo, D'Angelo, Di Costa, & Solazzi, 2011).

## 3. The assessment of the scholarly impact of research: a theoretical base

Generally speaking, the objective of research activity is to produce new knowledge. The new knowledge has a complex character of both tangible nature (different types of publications, patents, databases, etc.) and intangible nature (tacit knowledge, consulting activity, etc.). Large scale studies, such as ours, can only try to assess what is measurable, i.e. knowledge encoded in written form, and embed a number of assumptions and limitations. In their seminal work on assessing scientific research, Martin and Irvine (1983) state that "the impact of a publication describes its actual influence on surrounding research activities at a given time", and further on "it is the impact of a publication that is most closely linked to the notion of scientific progress - a paper creating a great impact represents a major contribution to knowledge at that time". Along the same lines, Moravcsik (1977) holds that the impact of a publication lies in its influence on subsequent publications, and that such influence will manifest itself by the influenced paper citing the influencing paper.

Scientometricians recur to citation-based metrics to assess the scholarly impact of research. The underlying rationale is that, in order for a research result (as any good and service) to have an impact, it has to be "used". The question is whether and to what extent citations are certifications of real use and representative of all uses. This point is critical. It is about whether one believes that the norm is that scientists cite papers to recognize their influence, being aware that exceptions (uncitedness, undercitation, and overcitation) occur (Mertonian or normative theory of citing – Merton, 1973), or that the opposite is true, i.e. that citing to give credit is the exception, while persuasion is the major motivation for citing (social constructivism – Mulkay, 1976; Bloor, 1976). Should the latter school of thought be right, the only other possible way to assess the scientific impact of publications would be by peer judgement, which is costly, time-consuming, subjective, and unsuitable for large-scale investigations.[1]

Citation-based measurement of impact embeds a tradeoff between the level of accuracy and timeliness in measurement. One of the critical issues in the reliability of citation indicators of impact concerns the rapidity with which citations accumulate:

---

[1] It has been shown that citation-based metrics have stronger predictive power of long-term impact than peer review (Abramo, D'Angelo, & Reale, 2019).

citations accrue with time, and no one can know for sure for how much time. Therefore, citation count can serve as a reliable proxy of the scholarly impact of a work if observed at a sufficient distance from the date of publication, or applying what is called a "citation time window" of adequate length. Yet, given an impact assessment aimed at informing policy and management decisions, no reasonable decision-maker could wait the necessary decades for completion of the citation life cycle.

In searching to improve the predictive power of early citations, scientometricians have proposed combining citation counts with other independent variables (Levitt & Thelwall, 2011; Stegehuis, Litvak, & Waltman, 2015). The latest contribution to the subject is by Abramo, D'Angelo, and Felici (2019). Differently from previous contributions in the literature, the authors provide: i) the weighted combinations of citations and IF, as a function of the citation time window and field of research, which best predict future impact; and ii) the predictive power of each combination.

## 4. Data and methods

A brief overview of the Italian research system will help the reader to better contextualize the empirical evidence. The main statistics on R&D expenditures position Italy among the bottom of the leading industrialized countries. In 2017, the Italian gross domestic expenditure on R&D (GERD) as a percentage of GDP was around 1.4 percent, as compared to 2.2 percent of average EU-28 (2.8 in the U.S.) (OECD, 2020a). GERD per capita population was 553 current PPP $ (855 in EU-28; 1687 in the U.S.). Basic research expenditure as a percentage of GDP was 0.30 (not available for EU-28; 0.47 in the U.S.). The percentage of GERD performed by the business enterprise sector was 62 (66 in EU-28; 73 in the U.S.). Total researchers per thousand total employment was (8.48 in EU-28; 9.23 in the U.S.).

On the scholarly output side, instead, Italy performs among the top countries. It has shown a steady growth in the world share of WoS indexed publications since 2001 (around 4 percent in 2017). This growth has not occurred to the detriment of the average impact of output, which has actually increased to about 1.4 (normalized) citations per paper, positioning Italy second to the UK in Europe (Consiglio Nazionale delle Ricerche, 2018). Italy ranks eighth in the world by number of publications, and fifth by number of citations. These results are partly ascribed to the introduction of performance-based funding and promotion schemes in the public sector.

The Italian industrial structure is characterized by a disproportionate number of micro and small enterprises and a few large corporations, mainly competing in low and middle-low-tech sectors (OECD, 2020b). Therefore, R&TD conducted in the private sector is definitely tilted towards technology development and applied research. Publications emerging from applied research in middle- and low-tech fields tend to be more responsive to local needs, and therefore less appealing to the world scholarly community, who generally prefer to engage in frontier research in emerging fields.

To answer the research questions, we will use a bibliometric approach based on the taxonomy shown in Table 1. Each publication is classified according to the affiliation of the authors and the type of collaboration. Along the first dimension we can have publications tagged as "Italian U&PROs only", "Italian Industry only", or "Italian Industry+U&PROs". We add the suffix Italian here (but we drop it from now onward, for simplicity) because in case of publications in collaboration with foreign organizations, we are not able to decipher the latter's legal nature. Thus "U&PROs only" means that the author(s) are Italian universities and/or public research organizations, but not Italian private corporations; the same is true for "Italian Industry only".

Along the second dimension, we name single-authored publications "Type 0". Those in collaboration can be either "Intramural" if all co-authors have the same affiliation, or extramural if they do not. The latter can be further distinguished in "Extramural - National only", if all co-authors belong to domestic organizations; "Extramural - International only" if all co-authors (of Italian publications) belong to foreign organizations only; and "Extramural - National + International" if the three or more organizations in the address list, at least two are domestic and at least one is foreign.

The bibliometric dataset is drawn from the Italian Observatory of Public Research (ORP), a database developed and maintained by Research Value, S.r.l. and derived under license from the WoS. Beginning from the raw data of Italian publications indexed in the WoS *core collection*, we apply a complex algorithm for unification of institutional affiliations. Such algorithm, mainly based on manual classification of all Italian addresses of U&PROs and private organizations, allows us to assign a specific collaboration type to each publication,[2] 721812 in all for the 2010-2017 period, as shown in Table 2. Publications by industry represent 3.6 percent of total, of which 2.5 percent in collaboration with U&PROs.

*Table 1: Collaboration types based on the authors' affiliations and nature of collaboration.*

| Collaboration type | U&PROs only | Industry + U&PROs | Industry only |
|---|---|---|---|
| None - Single author | Type 0_A | | Type 0_B |
| Intramural | Type 1 | | Type 7 |
| Extramural - National only | Type 2 | Type 5 | Type 8 |
| Extramural - International only | Type 3 | | Type 9 |
| Extramural - National + International | Type 4 | Type 6 | Type 10 |

*Table 2: Breakdown of the bibliometric dataset (2010-2017 Italian publications) by collaboration type*

| Collaboration type | U&PROs only | Industry + U&PROs | industry only | Total |
|---|---|---|---|---|
| Single author | 74031 (98.8%) | | 937 (1.2%) | 74968 (10.4%) |
| Intramural | 165028 (98.6%) | | 2407 (1.4%) | 167435 (23.2%) |
| Extramural -National | 167395 (92.9%) | 12692 (7.0%) | 184 (0.1%) | 180271 (25.0%) |
| Extramural - International | 180506 (97.8%) | | 4111 (2.2%) | 184617 (25.6%) |
| Extramural - National + International | 108827 (95.0%) | 5510 (4.8%) | 184 (0.2%) | 114521 (15.9%) |
| Total | 695787 (96.4%) | 18202 (2.5%) | 7823 (1.1%) | 721812 (100%) |

To assess possible impact differences across fields, we assign each publication to the SC of the hosting source and group the SCs into macro-areas.[3] Table 3 reports the breakdown of private sector publications by macro-area. Almost half of the 26025 publications authored by Industry fall in Engineering, followed by Physics (17.3%), Biology (15.1%), Biomedical Research and Clinical medicine (13.9% each). About 70% of private sector publications are in co-authorship with U&PROs. The macro-areas in which private-public sector collaboration is more intense are Psychology (83.1% of total publications in the macro-area are in co-authorship), Clinical Medicine (79.4%), Biology and Chemistry (76.8% each), Earth and space sciences (76.1%).

---

[2] Articles, reviews, conference proceedings, letters, books and book chapters.
[3] Our assignment of SCs to disciplinary areas (Mathematics; Physics; Chemistry; Earth and Space Sciences; Biology; Biomedical Research; Psychology; Clinical Medicine; Engineering; Economics; Law, political and social sciences) follows a pattern previously published in the ISI Journal Citation Reports website, although this information is no longer available through the Clarivate web portal.

*Table 3: Breakdown of Italian private sector 2010-2017 publications by macro-area*

| Macro-area | Industry only | Industry + U&PROs | Total |
|---|---|---|---|
| Engineering | 4305 (34.2%) | 8279 (65.8%) | 12584 (48.4%) |
| Physics | 1412 (31.4%) | 3078 (68.6%) | 4490 (17.3%) |
| Biology | 914 (23.2%) | 3028 (76.8%) | 3942 (15.1%) |
| Biomedical Research | 1111 (30.7%) | 2507 (69.3%) | 3618 (13.9%) |
| Clinical Medicine | 746 (20.6%) | 2869 (79.4%) | 3615 (13.9%) |
| Chemistry | 573 (23.2%) | 1896 (76.8%) | 2469 (9.5%) |
| Earth and Space Sciences | 566 (23.9%) | 1802 (76.1%) | 2368 (9.1%) |
| Economics | 296 (46.3%) | 344 (53.8%) | 640 (2.5%) |
| Law, political and social sciences | 249 (51.3%) | 236 (48.7%) | 485 (1.9%) |
| Mathematics | 114 (27.3%) | 304 (72.7%) | 418 (1.6%) |
| Art and Humanities | 96 (37.4%) | 161 (62.6%) | 257 (1.0%) |
| Multidisciplinary Sciences | 61 (29.8%) | 144 (70.2%) | 205 (0.8%) |
| Psychology | 13 (16.9%) | 64 (83.1%) | 77 (0.3%) |
| Total* | 7823 (30.1%) | 18202 (69.9%) | 26025 (100.0%) |

*\* The figure differs from the column total due to multiple counts for publications in two or more macro-areas.*

## 5. Results

In this section, we first provide descriptive statistics of two indicators: i) the average impact on scientific advancement of private-sector publications; and ii) the incidence of private sector highly cited articles (HCAs), both as compared to the public ones. HCAs are those falling in the top x% of the reference distribution, i.e. as compared to all WoS publications of the same year and subject category, for citations accrued.

Then, we will implement an OLS regression model to answer the research questions, while controlling for covariates which can affect citation rates. As for the measurement of impact, we will use a weighted combination of field-normalized citations (counted at 31/12/2019) and field-normalized IF (measured at year of publication) of the hosting journal, as proposed by Abramo, D'Angelo, and Felici (2019). This weighted combination is the one which best predicts the overall impact of a publication. Moreover, because both citations and IF are field normalized, we do not need to control for different citation rates across fields. Because of the skeweness of citation distributions, we apply a log transformation of the article impact and add one to allow for the inclusion of uncited publications ($Y = \ln(\text{Article impact} +1)$).

Table 4 shows the average impact and the incidence of HCAs for all publications in the dataset, by sector and collaboration type.

Industry only publications show an average impact lower than that recorded for U&PROs only publications, for all collaboration types (column 4 of Table 4). Values increase from 0.149 for single-author publications, to 0.188 for publications resulting from intramural collaborations, with a further increase to 0.215 for extramural national, and a maximum of 0.463 for international. A similar pattern can be observed for U&PROs only publications, for which the maximum (0.637) is actually recorded for publications resulting from both national and international extramural collaborations.

In other words, the scholarly impact of publications grows with the complexity of the collaboration featuring the publication byline, but it is systematically lower for private authored publications than for public ones.

The same conclusion can be reached considering the incidence of HCAs (columns 5-8 of Table 4). Only 0.1% of single-author private publications reach the status of Top1% for accrued citations, against 0.4% for single-author public publications. The comparison is always in favor of public publications, for any collaboration types and thresholds. 30.1% of total scientific public only publications deriving from extramural national and international collaboration is in the Top20% tail, vs 16.3% for private only publications, i.e. 51% higher and 19% lower than the expected value, respectively. The only exception is represented by the top1% publications resulting from extramural national collaboration: they represent 1.1% of the total private portfolio, against 0.4% of public.

*Table 4: Average impact and incidence of highly cited articles out for Industry and U&PROs Italian publications*

| Author | Collaboration type | Publications | Avg impact | Top1% | Top5% | Top10% | Top20% |
|---|---|---|---|---|---|---|---|
| Industry only | None - Single author | 937 | 0.149 | 0.1% | 0.9% | 2.1% | 7.2% |
| | Intramural | 2407 | 0.188 | 0.2% | 1.1% | 2.3% | 5.4% |
| | Extramural - National | 184 | 0.215 | 1.1% | 1.1% | 4.3% | 6.5% |
| | Extram. - International | 4111 | 0.463 | 1.0% | 5.3% | 10.4% | 20.5% |
| | Extram. - National+Intern. | 184 | 0.368 | 1.1% | 3.8% | 7.1% | 16.3% |
| Industry + U&PROs | Extram. - National | 12692 | 0.409 | 0.4% | 2.5% | 6.6% | 15.6% |
| | Extram. - National+Intern. | 5510 | 0.569 | 1.7% | 7.4% | 14.2% | 26.1% |
| U&PROs only | None - Single author | 74031 | 0.195 | 0.4% | 2.4% | 5.5% | 11.6% |
| | Intramural | 165028 | 0.365 | 0.4% | 2.6% | 6.0% | 13.9% |
| | Extramural - National | 167395 | 0.443 | 0.4% | 2.8% | 7.0% | 16.6% |
| | Extram. - International | 180506 | 0.587 | 1.8% | 7.9% | 14.7% | 27.7% |
| | Extram. - National+Intern. | 108827 | 0.637 | 2.0% | 8.7% | 16.3% | 30.1% |

In order to answer the research questions, while controlling for covariates which can affect citation rates, we will use an article level OLS regression model, specified from time to time according to the type of analysis to be carried out. Since, as shown above, statistics for the average impact and the incidence of HCAs are closely aligned, we will consider the former as a response variable.

We control for the number of authors in the byline, and for different document types through dummy variables (assuming as baseline the document type "article").

The combination of a set of other dummies allows for estimating the effect of various collaboration types (intramural, extramural, domestic, international) on impact of the resulting publication.

Estimation is performed with the vce(robust) option of Stata, which uses the sandwich estimator of variance, robust to some types of misspecification as long as the observations are independent.

The average values of the dependent variable and the covariates of the article level OLS regression model are shown in Table 5, by collaboration type. The number of authors of a publication varies according to the type of collaboration: for Industry only publications, the average value increases from 4.4 of intramural to 4.7 of national extramural, 7.5 of international and 7.9 of national+international collaborations.

A similar pattern can be observed for U&PROs only publications, for which the average value of 52.5 authors for extramural collaborations with both national and internatonal co-authors is due to outliers, publications characterized by mega-authorship (bylines with thousands of authors), typical of large research collaborations, especially in some fields of experimental physics.

As for document types, reviews represent less than 5% of total publications with relatively lower values for Industry only publications (2.2%-3.8%, depending on collaboration type) compared to U&PROs only (3.8%-8.3%). Conversely, conference proceedings represent a large share of Industry only publications (even 59.6% for intramural publications) while they are much less frequent among U&PROs only publications (26.9%).

*Table 5: Average values of variables in the OLS regression model*

| Author | Collaboration type | Obs. | Number of authors | Review | Proceeding | Other doc. types | Y |
|---|---|---|---|---|---|---|---|
| Industry only | None - Single author | 937 | 1 | 0.038 | 0.239 | 0.099 | 0.149 |
| | Intramural | 2407 | 4.4 | 0.031 | 0.596 | 0.006 | 0.188 |
| | Extramural - National | 184 | 4.7 | 0.038 | 0.359 | 0.000 | 0.215 |
| | Extram. - International | 4111 | 7.5 | 0.031 | 0.264 | 0.007 | 0.463 |
| | Extram. - National+Intern. | 184 | 7.9 | 0.022 | 0.310 | 0.103 | 0.368 |
| Industry + U&PROs | Extram. - National | 12692 | 6.3 | 0.031 | 0.222 | 0.003 | 0.409 |
| | Extram. - National+Intern. | 5510 | 12.5 | 0.039 | 0.156 | 0.008 | 0.569 |
| U&PROs only | None - Single author | 74031 | 1 | 0.038 | 0.106 | 0.148 | 0.195 |
| | Intramural | 165028 | 4.5 | 0.069 | 0.269 | 0.006 | 0.365 |
| | Extramural - National | 167395 | 6.4 | 0.083 | 0.063 | 0.004 | 0.443 |
| | Extram. - International | 180506 | 6.9 | 0.066 | 0.108 | 0.009 | 0.587 |
| | Extram. - National+Intern. | 108827 | 52.5 | 0.070 | 0.054 | 0.005 | 0.637 |

To answer the first research question, we use the following OLS regression model:

$$Y = b_0 + b_1 X_1 + b_2 X_2 + b_3 X_3 + b_4 X_4 + b_5 X_5 + b_6 X_6 + b_7 X_7$$

[1]

Where:

$Y = \ln(\text{Article impact} +1)$

$X_1$ = Number of authors (integer variable)

$X_2$ = Review (dummy)

$X_3$ = Proceeding (dummy)

$X_4$ = Other document types (dummy)

$X_5$ = International (dummy)

$X_6$ = Industry only (dummy)

$X_7$ = Industry + U&PROs (dummy)

In short, we check if, controlling for the number of authors, the internationalization of the byline, and the document type, Industry only publications have a different impact than U&PROs only.

The results shown in Table 6 indicate a negative effect on impact (-7.4%, i.e. $e^{-0.077}$)[4] for Industry only publications versus U&PROs only ones. The coefficient for Industry + U&PROs joint publications, on the other hand, is positive (+4.2%, i.e. $e^{0.041}$), meaning that collaboration with the private sector increases the impact of U&PROs publications, other things being equal. Note that publications with at least one foreign affiliation show a greater impact (+22.8, i.e. $e^{0.205}$). Not always the control variables do act in the expected direction: as for document type, reviews show a higher average impact (+27.0%, i.e. $e^{0.239}$) than articles, while proceedings and other document types show a lower impact (-26.7% and -31.2%, respectively). Differently from previous studies (Tahamtan, Safipour Afshar, & Ahamdzadeh, 2016), the variable "number of authors" shows a nihil coefficient, indicating the absence of effect on publication impact, when considering the other model covariates.

---

[4] Because of the log transformation of the impact, the marginal effect is represented by the exponential of the coefficient.

*Table 6: OLS regression for 2010-2017 Italian publications. Y=ln(Article impact +1)*

|  | Coeff. | Std Err. | t | P>\|t\| | [95% Conf. Interval] | |
|---|---|---|---|---|---|---|
| Const. | 0.409 | 0.001 | 595.0 | 0.000 | 0.407 | 0.410 |
| Number of authors | 0.000 | 0.000 | 28.9 | 0.000 | 0.000 | 0.000 |
| International | 0.205 | 0.001 | 185.7 | 0.000 | 0.203 | 0.207 |
| Industry only | -0.077 | 0.004 | -17.9 | 0.000 | -0.085 | -0.068 |
| Industry + U&PROs | 0.041 | 0.003 | 13.0 | 0.000 | 0.035 | 0.047 |
| Review | 0.239 | 0.003 | 89.6 | 0.000 | 0.234 | 0.244 |
| Proceeding | -0.310 | 0.001 | -291.6 | 0.000 | -0.312 | -0.308 |
| Other doc. types | -0.374 | 0.002 | -230.9 | 0.000 | -0.378 | -0.371 |

*Baselines: "U&PROs only", for publication type; "articles", for document types.*
*Number of obs. = 721812*
*F(7;721804) =27414.6; Prob > F = 0.000*
*R-squared = 0.146; Root MSE = 0.437*

In order to understand the effect of the type of collaboration on impact of publications, we apply an extended version of the OLS model:

$$Y = b_0 + b_1 X_1 + b_2 X_2 + b_3 X_3 + b_4 X_4 + b_5 X_5 + b_6 X_6 + b_7 X_7 + b_8 X_8$$

[2]

Where:

Y, $X_1$-$X_4$ as in [1]

$X_5$ = Intramural (dummy)

$X_6$ = Extramural - national (dummy)

$X_7$ = Extramural - international (dummy)

$X_8$ = Extramural - national + international (dummy)

The set of dummies $X_5$-$X_8$ represents the type of collaboration characterizing the publication, based on the relative byline and address list, assuming as baseline the "single-author paper" type.

Table 7 shows the results of the OLS regression applied to Industry only publications. The coefficient $b_5$, relative to the dummy "Intramural", equal to 0.069, indicates a positive and statistically significant effect of this type of collaboration on the impact of the resulting publications, compared to single-author ones. Then, with respect to single-author research projects, the collaborative work of researchers belonging to the same organization (intramural) has a greater impact, on average, of 7.1% ($e^{0.069}$). For publications resulting from national extramural collaboration, this effect is not significant, while the international openness of collaboration leads to a very evident increase in impact: +21.2% and +13.9% respectively, for extramural international and extramural both national and international.

*Table 7: OLS regression for Industry only 2010-2017 Italian publications. Y=ln(Article impact +1)*

|  |  | Coeff. | Std Err. | t | P>|t| | [95% Conf. Interval] | |
|---|---|---|---|---|---|---|---|
|  | Const. | 0.213 | 0.011 | 19.5 | 0.000 | 0.192 | 0.235 |
|  | Number of authors | 0.017 | 0.002 | 7.6 | 0.000 | 0.013 | 0.021 |
| Collab. type | Intramural | 0.069 | 0.014 | 4.9 | 0.000 | 0.041 | 0.096 |
| | Extramural - National | 0.016 | 0.026 | 0.6 | 0.545 | -0.035 | 0.066 |
| | Extram. - International | 0.192 | 0.018 | 11.0 | 0.000 | 0.158 | 0.226 |
| | Extram. - National+Intern. | 0.130 | 0.036 | 3.6 | 0.000 | 0.060 | 0.200 |
| Doc. type | Review | 0.357 | 0.037 | 9.7 | 0.000 | 0.285 | 0.429 |
| | Proceeding | -0.300 | 0.008 | -38.0 | 0.000 | -0.316 | -0.285 |
| | Other doc. types | -0.235 | 0.022 | -10.6 | 0.000 | -0.278 | -0.191 |

*Baselines: "single_author", for collaboration type; "articles", for document types.*
*Number of obs = 7823*
*F(8;7814) =374.43; Prob > F = 0.0000*
*R-squared = 0.296; Root MSE = 0.362*

Table 8 shows the comparison between the results of this OLS regression and those obtained by applying the same model to the dataset of U&PROs only publications. It can be noted that the increase in the impact of publications resulting from intramural collaborations compared to single-author papers has almost tripled for the latter set compared to the former (+19%, i.e. $e^{0.175}$ vs +7%, i.e. $e^{0.069}$). The increase in impact for national extramural collaborations for U&PROs only publications is in this case significant and rather relevant (+20%, i.e. $e^{0.184}$). Finally, the significant effect of international collaboration is confirmed and revealed to be higher for U&PROs only publications (+42% and +45%, i.e. $e^{0.348}$ and $e^{0.371}$ respectively) as compared to Industry only publications (+21% and +14%).

*Table 8: OLS regressions for 2010-2017 Italian publications. Y=ln(Article impact +1)*

|  |  | Industry only | U&PROs only |
|---|---|---|---|
|  | Const. | 0.213*** | 0.259*** |
|  | Number of authors | 0.017*** | 0.000*** |
| Collab. type | Intramural | 0.069*** | 0.175*** |
|  | Extramural - National | 0.016 | 0.184*** |
|  | Extram. - International | 0.192*** | 0.348*** |
|  | Extram. - National+Intern. | 0.130*** | 0.371*** |
| Doc. type | Review | 0.357*** | 0.231*** |
|  | Proceeding | -0.300*** | -0.310*** |
|  | Other doc. types | -0.235*** | -0.268*** |
|  | Obs | 7823 | 695787 |
|  | F | F(8;781)=374.43 | F(8;695778)=24843.35 |
|  | R-squared= | 0.296 | 0.156 |
|  | Root MSE= | 0.362 | 0.435 |

*Statistical significance: \* p-value <0.10, \*\* p-value <0.05, \*\*\* p-value <0.01.*

In order to understand how private-public collaboration affects the resulting publications' impact, in Table 9 we show the results of an OLS regression applied to all and only the publications resulting from national extramural collaboration. In particular, we consider a simplified model compared to [2], namely:

$$Y = c_0 + c_1 X_1 + c_2 X_2 + c_3 X_3 + c_4 X_4 + c_5 X_5$$

[3]

Where:

$X_5$ = dummy variable, assuming 1 if the publication is the outcome of a private-public joint collaboration; 0, otherwise

Y, $X_1$-$X_4$ as in [1].

This model will be applied, according to the taxonomy of Table 1, in a dual way to:
- the 12876 Type 5 and Type 8 publications, to measure the impact of collaboration with PROs, for Italian private companies;
- the 180087 Type 2 and Type 5 publications; to measure the impact of collaboration with private companies, for Italian PROs.

As for the first subset (upper part of Table 9), the coefficient $c_5$ (0.139) reveals a significant and quite high effect (+14.8%), indicating that in national extramural collaborations industry researchers benefit significantly from collaboration with colleagues affiliated to U&PROs, in terms of the impact of the resulting publications.

The dual analysis (in the lower part of Table 9), instead, shows a very small coefficient (0.023), although significant, to indicate that even for public researchers, collaborations with colleagues affiliated to private companies determines a positive impact differential (+2%, i.e. e0.023), but not so remarkable as for private researchers.

Finally, in Table 10 we show the results of the OLS regression [3] applied to the set of publications resulting from both national and international extramural collaboration. The upper part shows the results of the regression carried out on the set of publications authored by Industry (Type 6 + Type 10), the lower part on the set of publications authored by U&PROs (Type 4 + Type 6).

For private sector publications, the effect of collaboration with the public in publications including also a foreign author is positive. The coefficient of the covariate of interest indicates a marginal effect (+7%, i.e. e0.066) positive and significant on publications of this type, compared to those without the involvement of public co-authors.

On the contrary, for U&PROs publications the coefficient of the interest variable (-0.004) indicates a negative but not significant effect (p-value 0.530). In other words, in international research collaborations, U&PROs do not benefit from the co-occurrence of private Italian researchers also, at least in terms of impact of the resulting publications.

*Table 9: OLS regressions for 2010-2017 Extramural - national Italian publications. Y=ln(Article impact +1)*

| | | Coeff. | Std Err. | t | P>|t| | [95% Conf. Interval] | |
|---|---|---|---|---|---|---|---|
| **Industry only vs Industry+U&PROs** | Const. | 0.284 | 0.024 | 12.0 | 0.000 | 0.238 | 0.331 |
| | Number of authors | 0.008 | 0.001 | 7.9 | 0.000 | 0.006 | 0.010 |
| | Collaboration effect ($c_5$) | 0.139 | 0.023 | 6.0 | 0.000 | 0.094 | 0.184 |
| | Review | 0.250 | 0.027 | 9.2 | 0.000 | 0.197 | 0.303 |
| | Proceeding | -0.326 | 0.006 | -55.1 | 0.000 | -0.338 | -0.315 |
| | Other doc. types | -0.343 | 0.045 | -7.6 | 0.000 | -0.431 | -0.255 |
| | | | *Number of obs* | 12876 | | | |
| | | | *F(5;12870)* | 781.3 | | *Prob > F* | *0.000* |
| | | | *R-squared* | 0.148 | | *RootMSE* | *0.375* |
| **U&PROs only vs Industry+U&PROs** | Const. | 0.414 | 0.009 | 47.9 | 0.000 | 0.397 | 0.431 |
| | Number of authors | 0.005 | 0.001 | 3.9 | 0.000 | 0.003 | 0.008 |
| | Collaboration effect ($c_5$) | 0.023 | 0.004 | 6.4 | 0.000 | 0.016 | 0.030 |
| | Review | 0.183 | 0.005 | 40.7 | 0.000 | 0.174 | 0.192 |
| | Proceeding | -0.293 | 0.003 | -93.8 | 0.000 | -0.299 | -0.286 |
| | Other doc. types | -0.323 | 0.010 | -32.0 | 0.000 | -0.343 | -0.303 |
| | | | *Number of obs* | 180087 | | | |
| | | | *F(5;180081)* | 4180.9 | | *Prob > F* | *0.000* |
| | | | *R-squared* | 0.062 | | *RootMSE* | *0.389* |

*Table 10: OLS regressions for 2010-2017 Extramural - national and international Italian publications. Y=ln(Article impact +1)*

| | | Coeff. | Std Err. | t | P>|t| | [95% Conf. Interval] | |
|---|---|---|---|---|---|---|---|
| Industry only vs Industry +U&PROs | Const. | 0.519 | 0.033 | 15.9 | 0.000 | 0.455 | 0.584 |
| | Number of authors | 0.004 | 0.000 | 8.2 | 0.000 | 0.003 | 0.005 |
| | Collaboration effect ($c_5$) | 0.066 | 0.032 | 2.1 | 0.040 | 0.003 | 0.130 |
| | Review | 0.233 | 0.044 | 5.3 | 0.000 | 0.146 | 0.319 |
| | Proceeding | -0.447 | 0.012 | -38.6 | 0.000 | -0.470 | -0.425 |
| | Other doc. types | -0.464 | 0.047 | -9.8 | 0.000 | -0.557 | -0.371 |
| | | | *Number of obs* | 5694 | | | |
| | | | *F(5;5688)* | 368.9 | | *Prob > F* | *0.000* |
| | | | *R-squared* | 0.169 | | *RootMSE* | *0.482* |
| U&PROs only vs Industry+U&PROs | Const. | 0.639 | 0.002 | 377.5 | 0.000 | 0.636 | 0.642 |
| | Number of authors | 0.000 | 0.000 | 24.1 | 0.000 | 0.000 | 0.000 |
| | Collaboration effect ($c_5$) | -0.004 | 0.007 | -0.6 | 0.530 | -0.018 | 0.009 |
| | Review | 0.257 | 0.008 | 33.9 | 0.000 | 0.242 | 0.272 |
| | Proceeding | -0.474 | 0.004 | -125.9 | 0.000 | -0.482 | -0.467 |
| | Other doc. types | -0.495 | 0.013 | -37.2 | 0.000 | -0.521 | -0.469 |
| | | | *Number of obs* | 114337 | | | |
| | | | *F(5;114331)* | 4038.3 | | *Prob > F* | *0.000* |
| | | | *R-squared* | 0.073 | | *RootMSE* | *0.520* |

Given the small number of observations for private sector publications, an in-depth analysis by macro-area can only be conducted only for U&PROs publications. A first analysis concerns the effect of collaboration with Industry in national extramural collaborations. Table 11 reports the results of the OLS regression conducted according to the model [3], to investigate, at macro-area level, the presence of a difference in impact of publications authored by U&PROs only vs Industry+U&PROs joint publications. The $c_5$ coefficient value is significant in all areas but Biology, Law, political and social sciences, Art and Humanities. In the remaining macro-areas the sign is positive in Physics, Biomedical Research, Clinical Medicine and negative in the other six. Therefore, there are marginal effects, not always significant, not concordant in sign, and, in any case, very limited in amplitude, from -10% in Psychology to +3% in Biomedical Research.

Still referring to U&PROs publications but, in this case, resulting from extramural national and international collaboration, Table 12 indicates a significant effect of the collaboration with Industry in only 4 macro-areas. This is in any case a negative effect on the impact of the resulting publications and, more precisely: -11% in Mathematics; -7% in Earth and Space Sciences; -4% in Chemistry and Biomedical Research.

*Table 11: OLS regressions for 2010-2017 Extramural - national Italian U&PROs publications, by macro-area. Y=ln(Article impact +1)*

| Macro-area | Number of obs | Collaboration effect ($c_5$) | p-value | F | Prob > F | Root MSE | R-squared |
|---|---|---|---|---|---|---|---|
| Mathematics | 5556 | -0.076 | 0.005 | 377.8 | 0.000 | 0.432 | 0.080 |
| Physics | 20677 | 0.019 | 0.027 | 1571.9 | 0.000 | 0.390 | 0.146 |
| Chemistry | 14183 | -0.033 | 0.002 | 395.0 | 0.000 | 0.390 | 0.077 |
| Earth and Space Sciences | 12066 | -0.028 | 0.011 | 653.1 | 0.000 | 0.369 | 0.112 |
| Biology | 30068 | -0.012 | 0.169 | 795.9 | 0.000 | 0.376 | 0.071 |
| Biomedical Research | 35348 | 0.029 | 0.002 | 1073.7 | 0.000 | 0.376 | 0.052 |
| Clinical Medicine | 63923 | 0.024 | 0.007 | 798.7 | 0.000 | 0.368 | 0.049 |
| Psychology | 2997 | -0.110 | 0.055 | 119.9 | 0.000 | 0.373 | 0.061 |
| Engineering | 34252 | -0.040 | 0.000 | 2326.9 | 0.000 | 0.407 | 0.187 |
| Economics | 4273 | -0.064 | 0.011 | 178.2 | 0.000 | 0.406 | 0.099 |
| Law, political and social sciences | 4244 | -0.029 | 0.293 | 206.6 | 0.000 | 0.407 | 0.082 |
| Art and Humanities | 1774 | -0.043 | 0.232 | 58.7 | 0.000 | 0.398 | 0.090 |

*Table 12: OLS regressions for 2010-2017 Extramural - national and international Italian U&PROs publications, by macro-area. Y=ln(Article impact +1)*

| Macro-area | Number of obs | Collaboration effect ($c_5$) | p-value | F | Prob > F | Root MSE | R-squared |
|---|---|---|---|---|---|---|---|
| Mathematics | 2984 | -0.122 | 0.007 | 105.5 | 0.000 | 0.523 | 0.070 |
| Physics | 30408 | 0.015 | 0.349 | 2320.9 | 0.000 | 0.482 | 0.128 |
| Chemistry | 8258 | -0.037 | 0.071 | 132.0 | 0.000 | 0.461 | 0.069 |
| Earth and Space Sciences | 8010 | -0.072 | 0.000 | 431.0 | 0.000 | 0.459 | 0.130 |
| Biology | 16849 | 0.016 | 0.298 | 412.5 | 0.000 | 0.475 | 0.086 |
| Biomedical Research | 18377 | -0.037 | 0.044 | 1595.3 | 0.000 | 0.544 | 0.043 |
| Clinical Medicine | 32056 | -0.009 | 0.604 | 745.0 | 0.000 | 0.551 | 0.066 |
| Psychology | 1816 | -0.139 | 0.106 | 238.4 | 0.000 | 0.437 | 0.114 |
| Engineering | 18312 | -0.013 | 0.225 | 1088.9 | 0.000 | 0.480 | 0.168 |
| Economics | 2646 | -0.044 | 0.292 | 140.8 | 0.000 | 0.512 | 0.096 |
| Law, political and social sciences | 2146 | -0.046 | 0.421 | 146.4 | 0.000 | 0.504 | 0.163 |
| Art and Humanities | 700 | -0.020 | 0.778 | 25.3 | 0.000 | 0.476 | 0.123 |

## 6. Discussion and conclusions

The literature on the assessment of the scholarly impact of research is mainly focused on research performed at U&PROs. This work intends to add new knowledge to the little we know about the impact of private sector research on scientific advancement (Bloch, Ryan, & Andersen, 2019; Lebeau, Laframboise, Larivière, & Gingras, 2008).

Through the analysis of WoS indexed 2010-2017 Italian publications, we have found that 3.6 percent of total Italian publications are authored by the private sector. Almost 70% of such publications are in collaboration with U&PROs. Researchers from Industry differ from public colleagues in the document type chosen to encode the new scientific knowledge they produce: conference proceedings account for 25% of total private sector publications, compared to 13% of U&PROs ones. This reveals the preference of the private sector to attend conferences, which represent an opportunity to linger behind with U&PROs colleagues and exchange knowledge informally through personal contact. However, this also depends on the fact that almost 50% of total publications by the private sector falls in the engineering macro-area, where also for academics, regardless of collaboration with private colleagues, conference proceedings represent a very frequent way of encoding and diffusing the new knowledge produced.

Industry only publications show an average impact lower than that recorded for U&PROs only publications; similarly, the share of highly cited publications by the private sector is significantly lower than that by the public sector. Controlling for the number of authors, the level of internationalisation of the byline and document type, industry only publications show a 7.4% lower impact than their public counterparts.

The scientific impact of Canadian industry publications (not controlling for any covariates) has increased steadily since 1988 and, in 2005, was on par with that of university publications (Lebeau, Laframboise, Larivière, & Gingras, 2008). On the contrary, in the Danish case the impact of industry publications has been decreasing over time. In the period 2010-2013, it was 7.3% lower than national U&PROs publications and 32.7% lower than international U&PROs papers (Bloch, Ryan, & Andersen, 2019).

There are no enough studies from other countries to reach a robust conclusion that the underperformance of private scientific research is physiological, as can be expected given the different publication incentive systems at work in the private and public (publish or perish) sectors. As for the case of Italy, the results can be partly explained considering the structure of the industrial system, a system composed mostly by micro-, small- and medium-sized enterprises, specialized in middle-low and low tech sectors. In such sectors, private research tends to be geared to respond to local needs, therefore resulting less appealing to the international scientific community. Furthermore, prevalently small-sized enterprises are not likely to reach a critical mass of R&D spending to conduct high-quality scientific research, which in general is catch-up rather than frontier research. Finally, Italy ranks among the top countries in the world by average impact of publications, which might make the relative impact of private research worse that it would appear in absolute terms, if compared to that of other countries.

Coming to the effect of collaboration, for industry only publications, those stemming from intramural collaborations present an impact 7% higher than single-author publications. For U&PROs only publications it is 19% higher. This effect vanishes for intra-sector extramural domestic collaborations for publications by industry (maybe due to the low number of observations), while for U&PROs publications we recorded a further impact increase of 20%. The presence of foreign organizations in the research team seems to be correlated to impact, but the relevant effect for the private sector is less evident than that for U&PROs. Furthermore, the increase in in the number of organizations in the research team corresponds to an increase in the impact of the resulting publications, once again more evident for publications by U&PROs.

Public-private joint publications, in the absence of foreign partners, lead to publications of greater impact, remarkably compared to Industry only (+14.8%) and to a less extent with respect to those authored by scientists from national U&PROs only (+2%).

We can infer that, for public researchers, the incentive to collaborate with private companies exists but is extremely reduced compared to that observed for colleagues in the private sector. The analysis by macro-area reveals, however, that the positive effect is only visible in Physics, Biomedical Research, and Clinical Medicine. In other fields, it is either not significant or even negative.

In essence, it would seem that in public-private mixed teams it is the presence of public researchers to give the research results a greater relevance. It could be because of their higher contribution in terms of knowledge needed for solution of the faced problem, or because of their better skills and experience in writing research manuscripts and finding the right journals for submission. Looking at collaboration from the perspective of the public scientist, increasingly subject to assessments based on the impact of their scientific activity, all other motivations being equal, the collaboration with colleagues in the private sector does not seem so crucial. Considering the existence of transaction costs inherent in any collaboration, especially in cross-sector collaboration, this result highlights a trade-off between policies aimed at increasing public researchers' productivity and those aimed at encouraging their involvement in technology transfer activities, among which, joint research collaboration with industry is a relevant one.

This result partially confirms the conclusions of Bloch, Ryan, and Andersen (2019) who, observing the Danish scientific production, find little difference in citation impact for public-only and public-private collaborations. Our findings also align with Bloch et al., in confirming that the largest predictor of publication impact is the presence of international collaboration, both for private and for public sector publications.

Compared to the Danish study, a very different conclusion emerges regarding the effect of public-private collaboration in internationally co-authored publications. While Bloch, Ryan, and Andersen (2019) found a much higher citation impact for such papers than public only publications, for the Italian case, the impact of publications resulting from international collaboration of U&PROs is negatively related to the presence of private sector co-authors. This is statistically evident both at overall level and in four specific macro-areas and more precisely in Mathematics, Earth and Space Sciences, Chemistry and Biomedical Research. These differences might be partly explained by the structure of the industrial system in the two countries, the Danish one being much smaller but featured by a very significant presence of large global corporations.

Finally, we remark the value of a byproduct of our work, of interest for the literature on the determinants of the impact of a publication emerges: among the factors affecting the number of citations received by a publication, contrary to what claimed by other scholars, the number of authors in the byline seems to have no effect when considering other covariates associated to the byline itself, including the type of collaboration (international vs national; extramural vs intramural). This reinforces what also Bloch, Ryan, and Andersen (2019) stated saying that "The number of co-authors does not fully explain differences in citation impact across types of collaboration".

In concluding this study, we cannot help reminding the reader of the main limits embedded in our analyses. First of all, our statistical model assumes that team composition has a causal impact on the outcomes of research activities. In practice, team composition and research agendas tend to be reciprocally determinated, and both are affected by a set of additional factors (e.g. team leader's network and experience, funding conditions, etc.) that should be controlled for. Because of the lack of data in practice, this is not possible at the level of individual publications for large-scale datasets. Moreover, all assumptions and limits of the bibliometric analysis apply. First, the new knowledge produced is not only those embedded in publications, and the bibliographic repertories (such as WoS, used here) do not register all publications. Furthermore, the measurement of the value of publications using citation-based indicators is a prediction, not definitive. Finally, citations can also be negative or inappropriate, and in any case they certify only scholarly impact, forgoing other types of impact.

These limitations are particularly relevant when the analysis refers to private R&D and should induce caution in interpreting findings arising from bibliometric techniques.